\documentclass[12pt]{iopart}

\usepackage{iopams}  
\usepackage{color}
\usepackage{graphicx}
\begin{document}

\title{Helicon plasma in a magnetic shuttle}
\author{L. Chang$^{1}$, J. Liu$^{2, 3}$, X. G. Yuan$^{1, 4}$, X. Yang$^1$, H. S. Zhou$^{1*}$, G. N. Luo$^1$, X. J. Zhang$^1$, Y. K. Peng$^1$, J. Dai$^{2, 3}$, G. R. Hang$^{2, 3}$}
\address{$^1$Institute of Plasma Physics, Chinese Academy of Sciences, Hefei 230031, China}
\address{$^2$Shanghai Institute of Space Propulsion, Shanghai 201112, China}
\address{$^3$Shanghai Engineering Research Center of Space Engine, Shanghai 201112, China}
\address{$^4$Graduate Island of Sciences, University of Science and Technology of China, Hefei 230026, China}
\ead{haishanzhou@ipp.ac.cn}

\date{\today}

\begin{abstract}
The definition of magnetic shuttle is introduced to describe the magnetic space enclosed by two tandem magnetic mirrors with the same field direction and high mirror ratio. Helicon plasma immersed in such a magnetic shuttle which can provide the confinement of charged particles is modeled using an electromagnetic solver. The perpendicular structure of wave field along this shuttle is given in terms of stream vector plots, showing significant change from midplane to ending throats, and the vector field rotates and forms a circular layer that separating plasma column radially into core and edge regions near the throats. The influences of driving frequency, plasma density and field strength on the wave field and power absorption are computed in detail. It is found that the wave magnitude and power absorption decrease for increased driving frequency and reduced field strength, and maximize around a certain level of plasma density. The axial standing-wave feature always exists, due to the interference between forward and reflected waves from ending magnetic mirrors, while the radial wave field structure largely stays the same. Distributions of wave energy density and power absorption density all show shrinking feature from midplane to ending throats, which is consistent with the nature of helicon mode that propagating along field lines. Theoretical analysis based on a simple magnetic shuttle and the governing equation of helicon waves shows consistency with computed results and previous studies. 
\end{abstract}



\maketitle

\section{Background}\label{bkg}
Helicon plasma has been finding emerging applications in material processing\cite{Lieberman:2005aa, Blackwell:2012aa, Rapp:2016aa}, space propulsion\cite{Chang-Diaz:2000aa, Charles:2002aa, Ziemba:2005aa, Batishchev:2009aa, Shinohara:2014aa} and the fundamental study of plasma physics\cite{Loewenhardt:1991aa, Petrzilka:1994aa, Hanna:2001aa, Chang:2011aa, Chang:2013aa}, due to its remarkable ionization efficiency and high plasma density. Although most studies treat helicon plasma in uniform magnetic field, the field geometry is eventually non-uniform, for example closed curves, because of its divergence-free nature (absence of magnetic charge). This applies to the natural magnetosphere around earth and laboratory solenoid coils usually employed for magnetic field generation. Previous attention has been given to non-uniform geometries such as cusped\cite{Harada:2014aa, Stenzel:2018aa}, focused\cite{Blackwell:2012aa, Stenzel:2018aa, Yoshitaka:2006aa}, diverged\cite{Lafleur:2011aa, Robertson:2017aa, Ghosh:2017aa} and rippled\cite{Chang:2013aa, Chang:2014aa} magnetic fields, however, little attention was given to multiple magnetic mirrors\cite{Takeno:1995aa, Belov:2007aa, Sahu:2013aa}. Here, we define ``magnetic shuttle" as two tandem magnetic mirrors which have the same field direction and high mirror ratio, and enclose a shuttle-shaped magnetic space. Please note that a magnetic mirror only refers to one single reflecting end, according to its physical meaning\cite{Chen:1984aa}, while the magnetic shuttle comprises two magnetic mirrors. This concept is also different from the same phrase (but different meaning) in quantum mechanics which describes the shuttling movement of electrons and involves metallic electrodes\cite{Kulinich:2014aa, Ilinskaya:2018aa, Ilinskaya:2019aa}. The magnetic shuttle could provide the confinement of charged particles due to Lorentz force and  the invariance of magnetic moment, and enables helicon plasma to capture the particle confinement physics relevant to fusion and space plasmas. The Helicon Physics Prototype eXperiment (HPPX\cite{Ping:2019aa}) constructed at Institute of Plasma Physics, Chinese Academy of Sciences, utilizes such a magnetic shuttle to produce high-density and large-volume helicon plasma and confine it thereafter. This work is devoted to analyzing the general physics of helicon plasma in a magnetic shuttle, and computing the wave physics and power absorption based on the real parameters of HPPX to provide guidance for ongoing experiments. 

\section{Electromagnetic solver}\label{ems}
\subsection{Governing equations}\label{eqn}
We employ an electromagnetic solver (EMS\cite{Chen:2006aa}) to compute the wave field and power absorption in helicon plasma which is confined by a magnetic shuttle. The solver has been used successfully in modeling various helicon sources, such as the helicon discharge machine at University of Texas at Austin\cite{Lee:2011aa}, the LArge Plasma Device (LAPD\cite{Gekelman:1991aa}) in University of California at Los Angles\cite{Chang:2014aa, Zhang:2008aa, Chang:2016aa, Chang:2018aa}, and the MAGnetized Plasma Interaction Experiment (MAGPIE\cite{Blackwell:2012aa}) at Australian National University\cite{Chang:2012aa}. It utilizes two Maxwell's equations to determine the wave field in helicon plasma: 
\begin{equation}\label{eq1}
\bigtriangledown\times\mathbf{E}=i \omega\mathbf{B}, 
\end{equation}
\begin{equation}\label{eq2}
\frac{1}{\mu_0}\bigtriangledown\times\mathbf{B}=-i \omega \mathbf{D}+\mathbf{j_a}.
\end{equation}
Here, $\mathbf{E}$ and $\mathbf{B}$ are wave electric and magnetic fields, respectively, $\mu_0$ is the permeability of vacuum, and $\mathbf{j_a}$ is the current density of external driving antenna. Perturbations in form of $\textrm{exp}[i(k z+m\theta-\omega t)]$ and right-hand cylindrical coordinate system $(r;~\theta;~z)$ are considered, with $k$ and $m$ the axial and azimuthal mode numbers, respectively, and $\omega$ the wave frequency, which is same to the frequency of external antenna for this driven system. A cold-plasma dielectric tensor is employed to link electric displacement vector $\mathbf{D}$ and $\mathbf{E}$ in form of
\begin{equation}\label{eq3}
\mathbf{D}=\varepsilon_0[\varepsilon\mathbf{E}+i g(\mathbf{E}\times\mathbf{b})+(\eta-\varepsilon)(\mathbf{E}\cdot\mathbf{b})\mathbf{b}],
\end{equation}
with $\varepsilon_0$ the permittivity of vacuum and $b$ the unit vector of guiding magnetic field. The components of dielectric tensor can be expressed as\cite{Ginzburg:1970aa}:
\begin{equation}\label{eq4}
\begin{array}{l}
\vspace{0.3cm}\varepsilon=1-\sum\limits_{\alpha}\frac{\omega+i\nu_\alpha}{\omega}\frac{\omega^2_{p \alpha}}{(\omega+i\nu_\alpha)^2-\omega^2_{c\alpha}},\\
\vspace{0.3cm}g=-\sum\limits_{\alpha}\frac{\omega_{c \alpha}}{\omega}\frac{\omega^2_{p\alpha}}{(\omega+i\nu_\alpha)^2-\omega^2_{c\alpha}},\\
\eta=1-\sum\limits_{\alpha}\frac{\omega^2_{p\alpha}}{\omega(\omega+i\nu_\alpha)}.
\end{array}
\end{equation}
While the subscript $\alpha$ denotes the different species of particles, namely ion and electron, the parameters $\omega_{p\alpha}$ and $\omega_{c\alpha}$ are plasma frequency and cyclotron frequency, respectively. Here, a phenomenological collision frequency $\nu_{\alpha}$ has been introduced. For typical helicon discharges, electron-neutral collision frequency is an order of magnitude lower than electron-ion collision frequency, and the relationship of $\nu_{ie}\ll\nu_{ei}$ usually applies\cite{Chang:2012aa}. Hence, we consider $\nu_{\alpha}\approx\nu_{ei}$ throughout the manuscript. 

We assume the equilibrium magnetic field ($B_0$) is axisymmetric, namely $B_{0r}~\ll~B_{0z}$ and $B_{0\theta}=0$, so as to make use of near axis expansion
\begin{equation}\label{eq5}
B_{0r}(r,~z)=-\frac{1}{2}r\frac{\partial B_{0z} (z)}{\partial z}. 
\end{equation}
For the right-hand half-turn helical antenna employed, we assume the current density is divergence free to eliminate capacitive coupling effects, and use its Fourier components of the form:
\begin{equation}\label{eq6}
\begin{array}{l}
\vspace{0.5cm}j_{ar}=0, \\
\vspace{0.3cm}j_{a\theta}=I_a\frac{e^{i m \pi}-1}{2}\delta(r-R_a)\left\{\frac{i}{m \pi}[\delta(z-z_a)+\delta(z-z_a-L_a)]\right.\\
\vspace{0.5cm}\hspace{0.9cm}\left.+\frac{H(z-z_a)H(z_a+L_a-z)}{L_a}e^{-i m \pi[1-(z-z_a)/L_a]}\right\}, \\
\vspace{0.3cm}j_{az}=I_a\frac{e^{-i m \pi[1-(z-z_a)/L_a]}}{\pi R_a}\frac{1-e^{i m \pi}}{2}\delta(r-R_a)\\
\hspace{0.9cm}\times H(z-z_a)H(z_a+L_a-z). \\
\end{array}
\end{equation}
With the subscript $a$ labeling antenna, $L_a$ and $R_a$ are length and radius, respectively, $z_a$ is the distance to left endplate, and $H$ is the Heaviside step function. 

\subsection{Computational implementations}\label{cmp}
\begin{figure}[ht]
\begin{center}$
\begin{array}{l}
(a)\\
\hspace{0.98cm}\includegraphics[width=0.63\textwidth,angle=0]{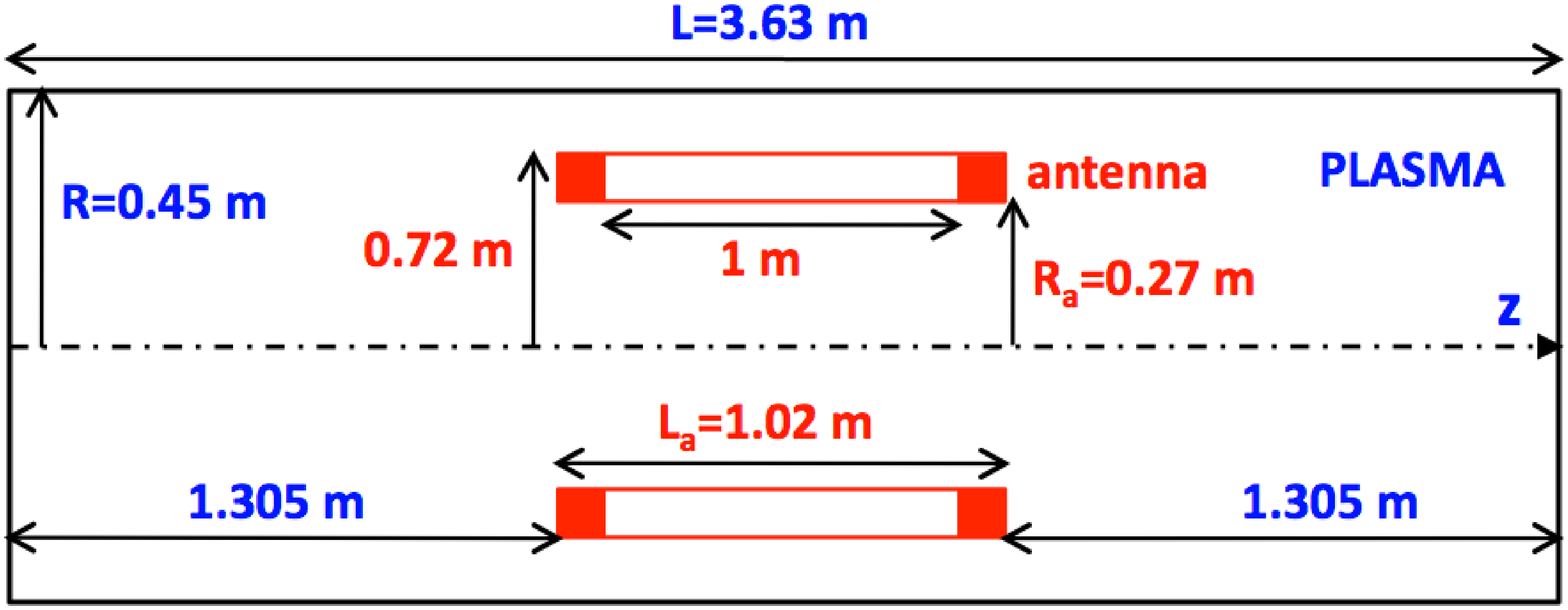}\\
(b)\\
\includegraphics[width=0.69\textwidth,angle=0]{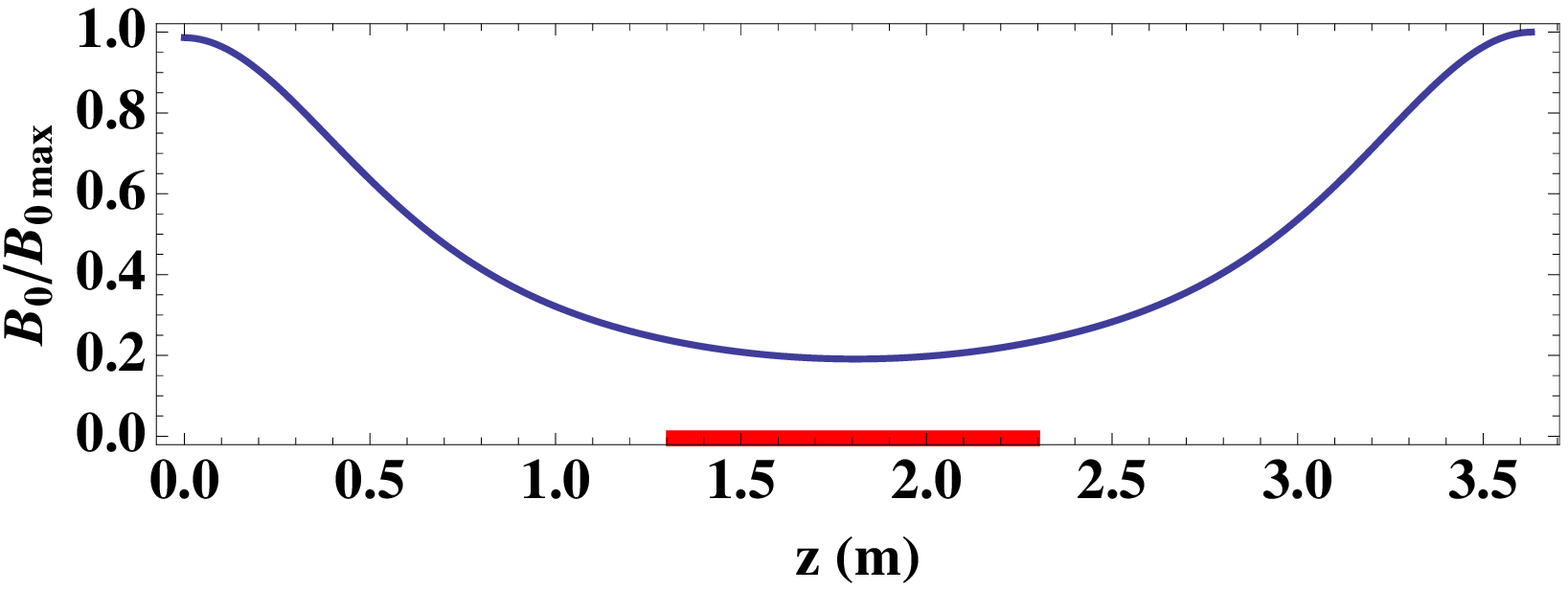}
\end{array}$
\end{center}
\caption{Employed computational domain and equilibrium magnetic field (normalized to its maximum value). Red bar shows the half-turn helical antenna.}
\label{fg1}
\end{figure}

The model described above is solved in a computational domain shown in Fig.~\ref{fg1}(a), based on a finite difference scheme. 
The inner length and radius of discharge chamber are $L=3.63$~m and $R=0.45$~m, respectively, while for the antenna they become $L_a=1$~m and $R_a=0.27$~m. To compare the upstream and downstream wave physics and power absorption, the antenna is located axially in the middle. For ideally conducting boundary conditions, the tangential components of $\mathbf{E}$ vanish at walls:
\begin{equation}\label{eq7}
\begin{array}{l}
\vspace{0.3cm}E_\theta(R, z)=E_z(R, z)=0, \\
\vspace{0.3cm}E_r(r, 0)=E_\theta(r, 0)=0, \\
E_r(r, L)=E_\theta(r, L)=0.
\end{array}
\end{equation}
Additionally, all field components must be regular on axis, namely $B_\theta |_{r=0}=0$ and $(rE_\theta)|_{r=0}=0$ for $m=~0$; $E_z|_{r=0}=0$ and $(rE_\theta)|_{r=0}=0$ for $m\neq 0$\cite{Zhang:2008aa}. In present computations, we set $m=1$ that is preferentially excited by half-turn helical antenna in helicon plasma\cite{Chen:1996aa, Light:1995aa, Light:1995ab}. 

The axial profile of equilibrium magnetic field is given in Fig.~\ref{fg1}(b), which has been normalized to its maximum value. Here, we choose three orders of field strength for comparison: $B_{0max}=0.017$~T, $B_{0max}=0.17$~T, $B_{0max}=1.7$~T. The depth of magnetic mirror is kept the same: $1-B_{0min}/B_{0max}=0.8$, which is quite large compared to typical experiments\cite{Zhang:2008aa} and enough for the confinement demonstration. The employed plasma density is in form of $n_e(r)=n_{emax}\textrm{exp}(-75.2 r^2)$ and does not have axial structure. It has been shrunk radially towards axis to lower the density level near the antenna and wall, which could help eliminate sparks according to previous experiments\cite{Loewenhardt:1991aa}. Four orders of plasma density are considered: $n_{emax}=10^{16}~\textrm{m}^{-3}$, $n_{emax}=10^{17}~\textrm{m}^{-3}$, $n_{emax}=10^{18}~\textrm{m}^{-3}$, $n_{emax}=10^{19}~\textrm{m}^{-3}$, referring to typical helicon discharges\cite{Blackwell:2012aa, Boswell:1970aa, Boswell:1987aa, Chen:1991aa, Chen:1996ab, Shinohara:2018aa}. Other input parameters include electron temperature of $T_e=8$~eV, helium, antenna current of $4.3$~A (referring to the MAGPIE experiment\cite{Blackwell:2012aa, Chang:2012aa}), and driving frequency ($f=\omega/2\pi$) ranging from $6.68$~MHz to $40.58$~MHz which also covers most helicon discharges. 

\section{Wave field and power absorption}\label{rst}
\subsection{Wave field structure along magnetic shuttle}\label{wfd}
\begin{figure}[ht]
\begin{center}
\includegraphics[width=1\textwidth,angle=0]{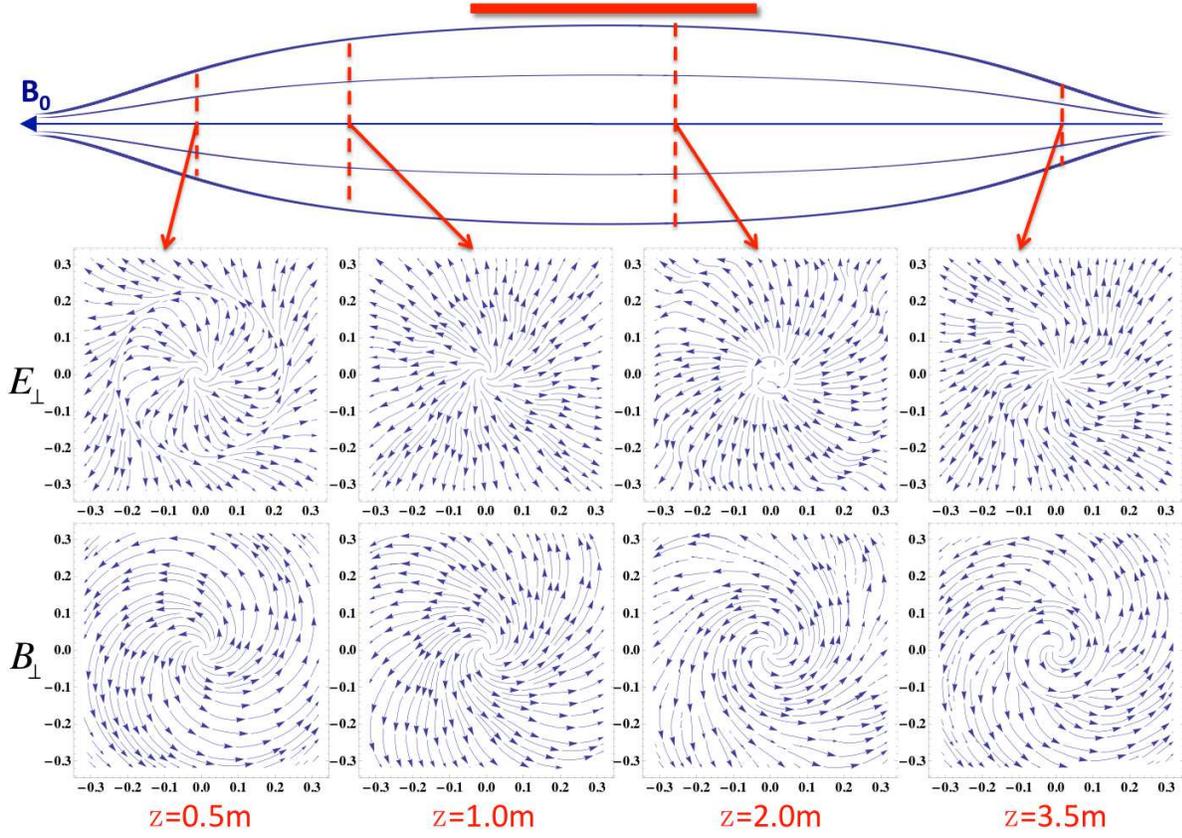}
\end{center}
\caption{Perpendicular wave field structure for $n_{emax}=10^{18}~\textrm{m}^{-3}$, $B_{0max}=0.17~\textrm{T}$ and $f=13.46$~MHz at different axial locations. Red bar labels the location of antenna.}
\label{fg2}
\end{figure}

The perpendicular wave field structure is shown in Fig.~\ref{fg2} in form of 2-dimensional (2D) stream vectors across plasma column. Please note that the field lines forming magnetic shuttle above are taken from Fig.~\ref{fg1}(b) with different scalings, hence they are physically meaningful. The employed conditions include $B_{0max}=0.17$~T, $n_{emax}=10^{18}~\textrm{m}^{-3}$ and $f=13.46$~MHz, which are typical levels for helicon discharge. The plots are measured at $0.5$~m, $1$~m, $2$~m, $3.5$~m, respectively, stepping from left endplate to right endplate through the whole magnetic shuttle.  

It can be seen that both the electric and magnetic wave field structures vary greatly with axial location, especially when approaching the ending magnetic throats: the vector field rotates and forms a circular layer that separates the plasma column radially into core and edge regions. Moreover, the presence of antenna also effects the wave field structure. Actually, it is the combination of reflection from ending throats and near field interference from antenna that makes the wave field profiles complicated (see Fig.~\ref{fg3}, Fig.~\ref{fg5}, Fig.~\ref{fg7} in following sections).

\subsection{Driving-frequency dependence}\label{frq}
\begin{figure}[ht]
\begin{center}$
\begin{array}{ll}
(a)&(b)\\
\includegraphics[width=0.45\textwidth,angle=0]{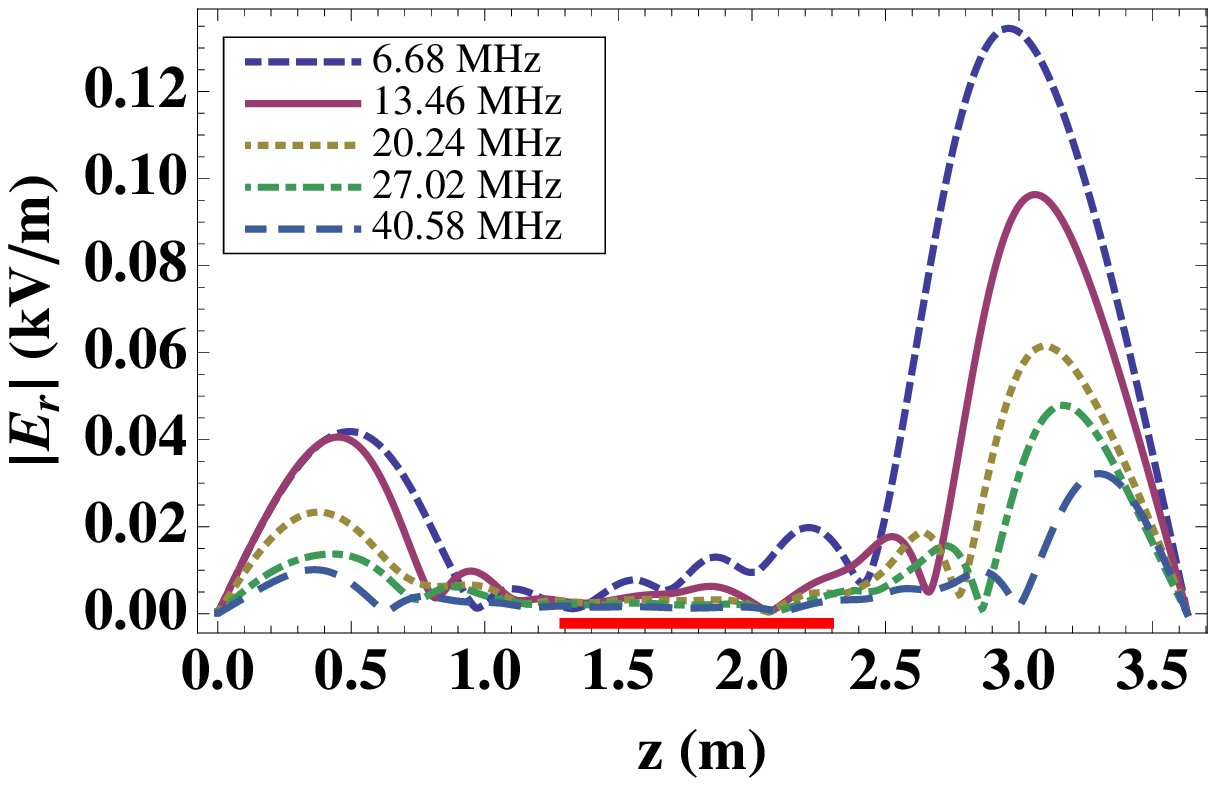}&\includegraphics[width=0.45\textwidth,angle=0]{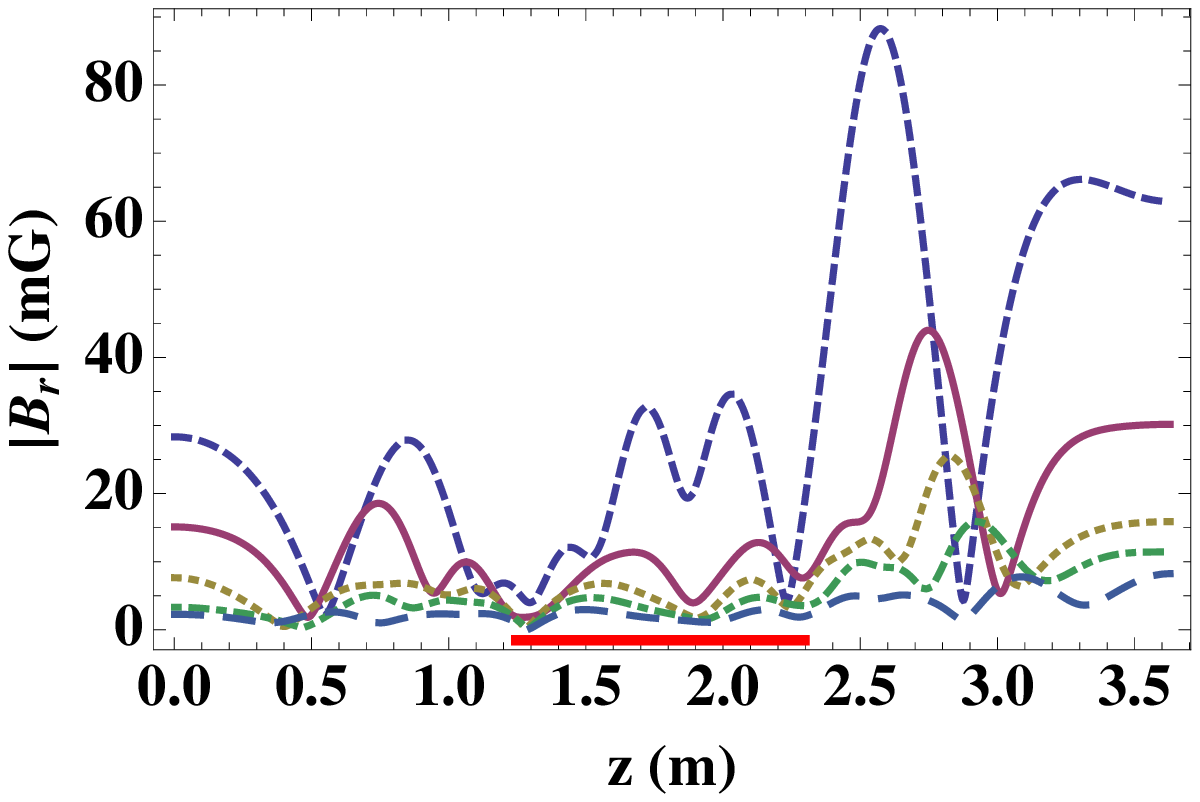}\\
\end{array}$
\end{center}
\caption{Variation of axial wave field profile ($r=0$~m) with driving frequency for $n_{emax}=10^{18}~\textrm{m}^{-3}$ and $B_{0max}=0.17~\textrm{T}$. Red bar shows the location of antenna.}
\label{fg3}
\end{figure}

To see the influence of driving frequency, the most controllable parameter in experiment, we compute the wave field for frequency range of $6.68\sim 40.58$~MHz. Other conditions are kept the same as for Fig.~\ref{fg2}. Typical results for $6.68$~MHz, $13.46$~MHz, $20.24$~MHz, $27.02$~MHz and $40.58$~MHz are shown in Fig.~\ref{fg3}. Please note that for the $m=1$ mode considered here, the azimuthal component is similar to the radial component, and the axial component is relatively much smaller. Therefore, only the radial components of wave electric and magnetic field are presented. 

From the axial profiles measured at $r=0$~m, we can see that the wave magnitude decreases for increased driving frequency, and the location of peak magnitude moves closer to ending throats. Importantly, these axial profile all show standing-wave feature which is caused by the interference between forward and reflected waves from ending throats. The asymmetrical feature of larger wave magnitude on the right-hand side of antenna than that on the left-hand side is caused by the directional preference of helical antenna, consistent with previous studies\cite{Lee:2011aa, Chang:2012aa, Chen:1996aa}. The radial profiles are largely the same for different driving frequencies.

\begin{figure}[ht]
\begin{center}$
\begin{array}{ll}
(a)&(b)\\
\includegraphics[width=0.48\textwidth,angle=0]{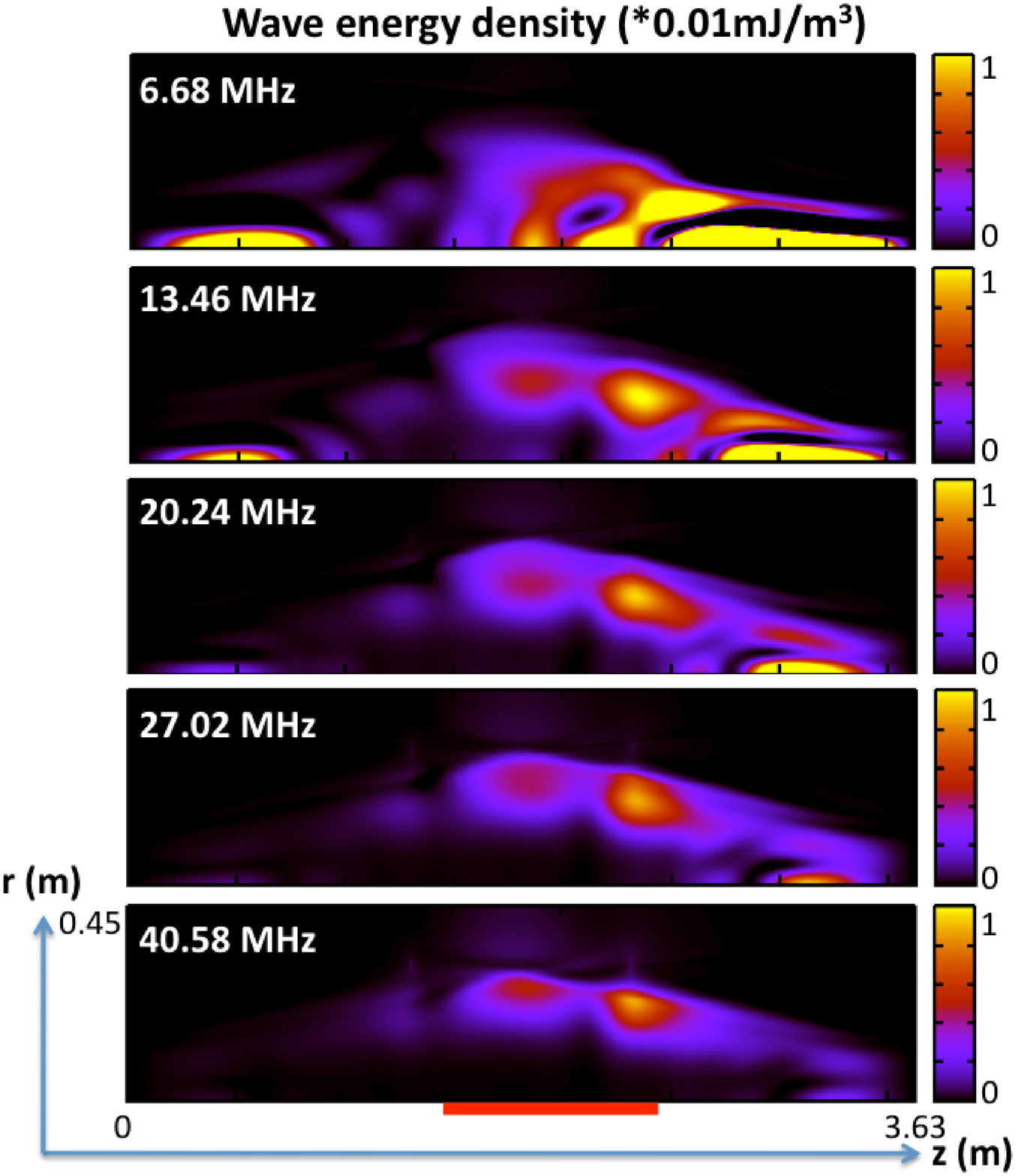}&\includegraphics[width=0.485\textwidth,angle=0]{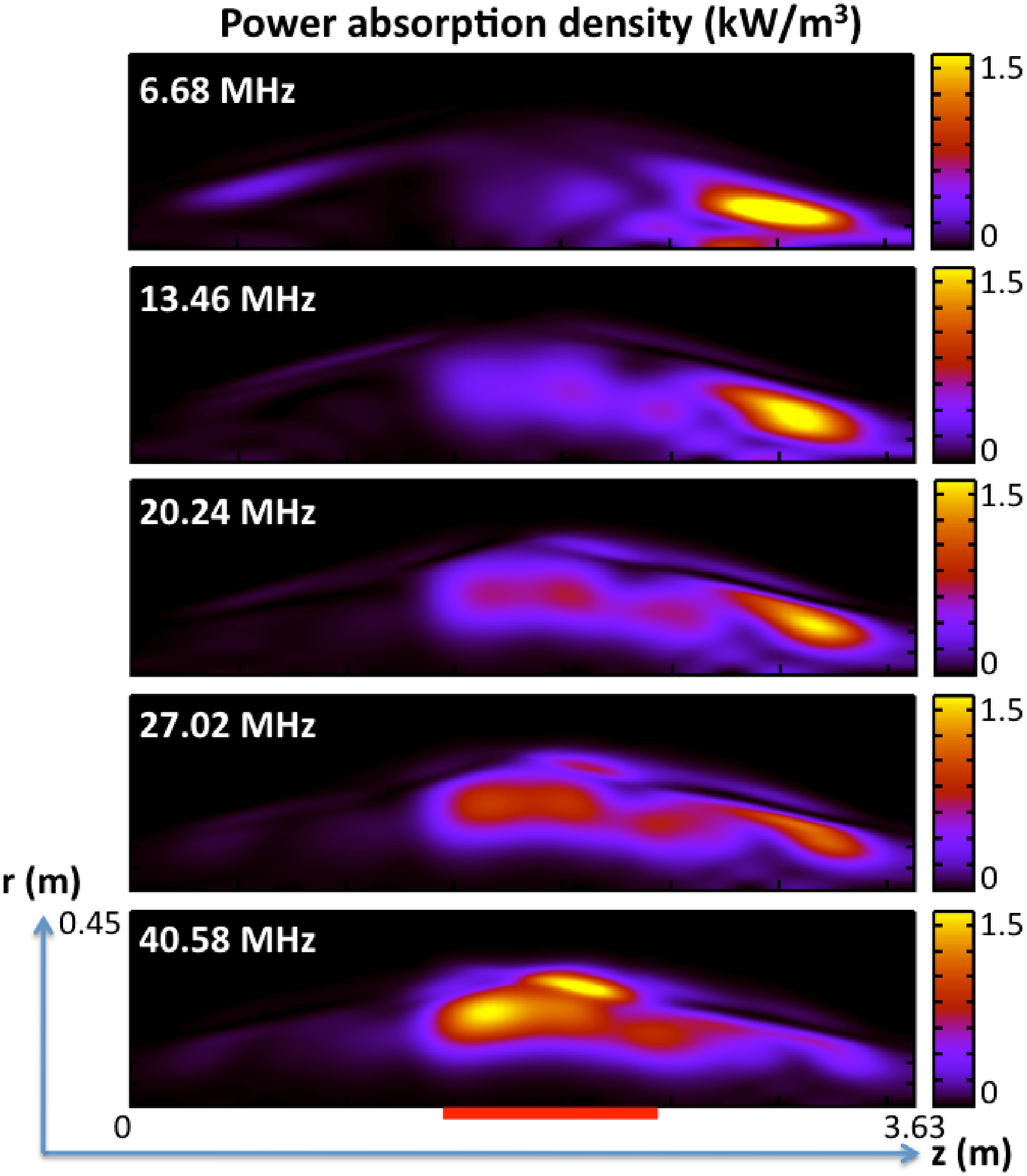}
\end{array}$
\end{center}
\caption{Variations of wave energy density (a) and power absorption density (b) with driving frequency for $n_{emax}=10^{18}~\textrm{m}^{-3}$ and $B_{0max}=0.17~\textrm{T}$. Red bar shows the location of antenna.}
\label{fg4}
\end{figure}

The 2D contour plots in Fig.~\ref{fg4} also show this asymmetrical feature: wave energy density and power absorption density are higher on the right-hand of antenna, and they largely decrease and localize more closely to antenna for increased driving frequency. Moreover, these contour plots display shrinking feature when looking from antenna to ending throats, which is another characteristic of magnetic mirror, besides the standing wave field structure. 

\subsection{Plasma-density dependence}\label{dst}
\begin{figure}[ht]
\begin{center}$
\begin{array}{ll}
(a)&(b)\\
\includegraphics[width=0.45\textwidth,angle=0]{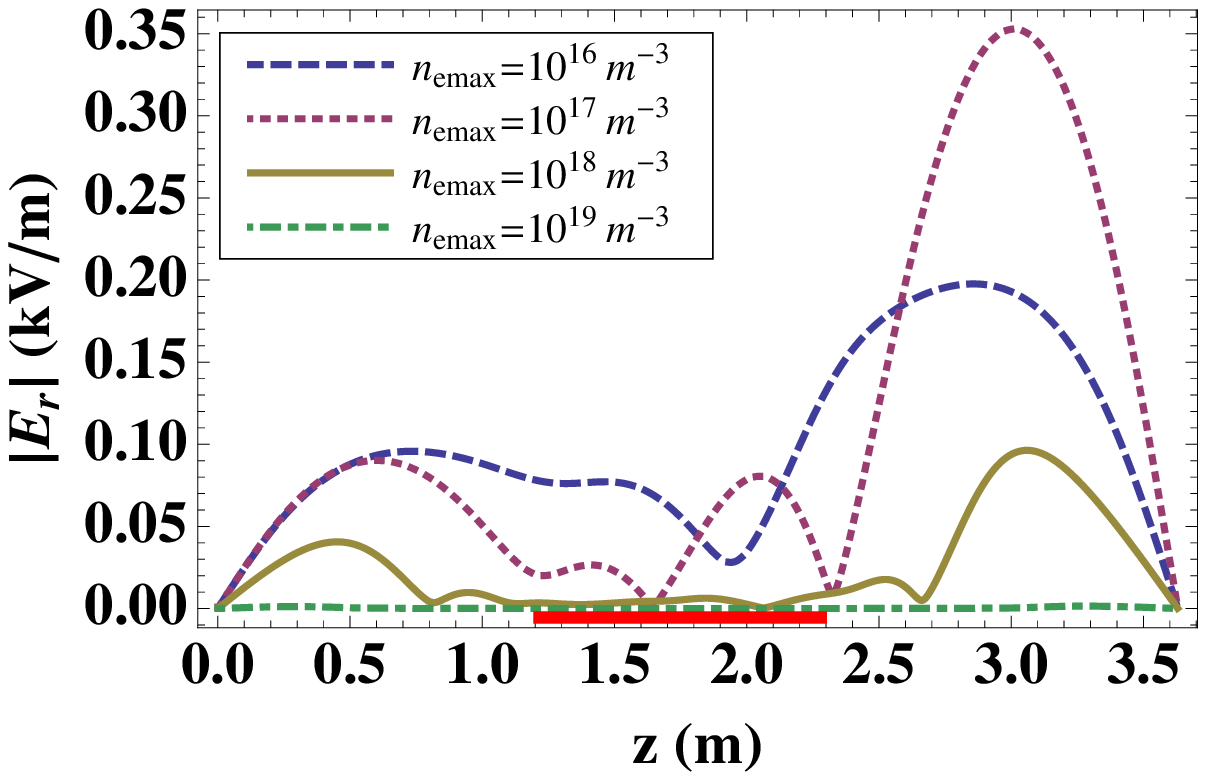}&\includegraphics[width=0.45\textwidth,angle=0]{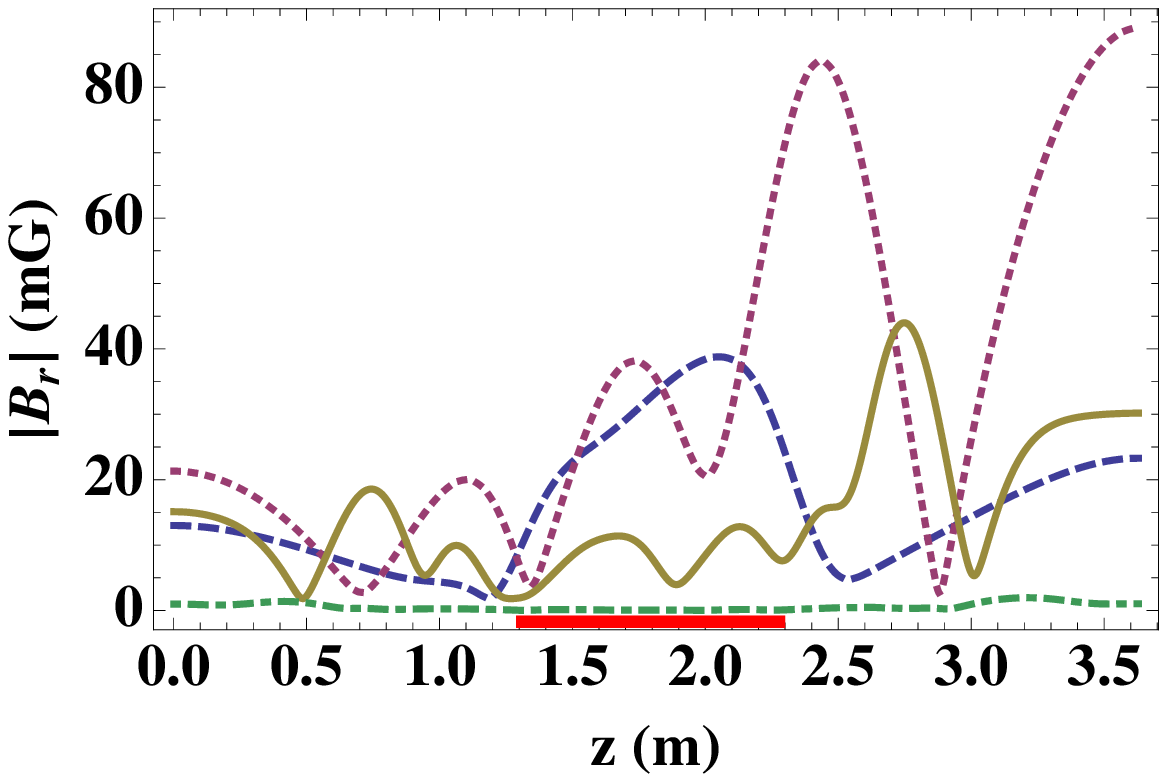}\\
\end{array}$
\end{center}
\caption{Variation of axial wave field profile ($r=0$~m) with plasma density for $f=13.46$~MHz and $B_{0max}=0.17~\textrm{T}$. Red bar shows the antenna location.}
\label{fg5}
\end{figure}

\begin{figure}[ht]
\begin{center}$
\begin{array}{ll}
(a)&(b)\\
\includegraphics[width=0.485\textwidth,angle=0]{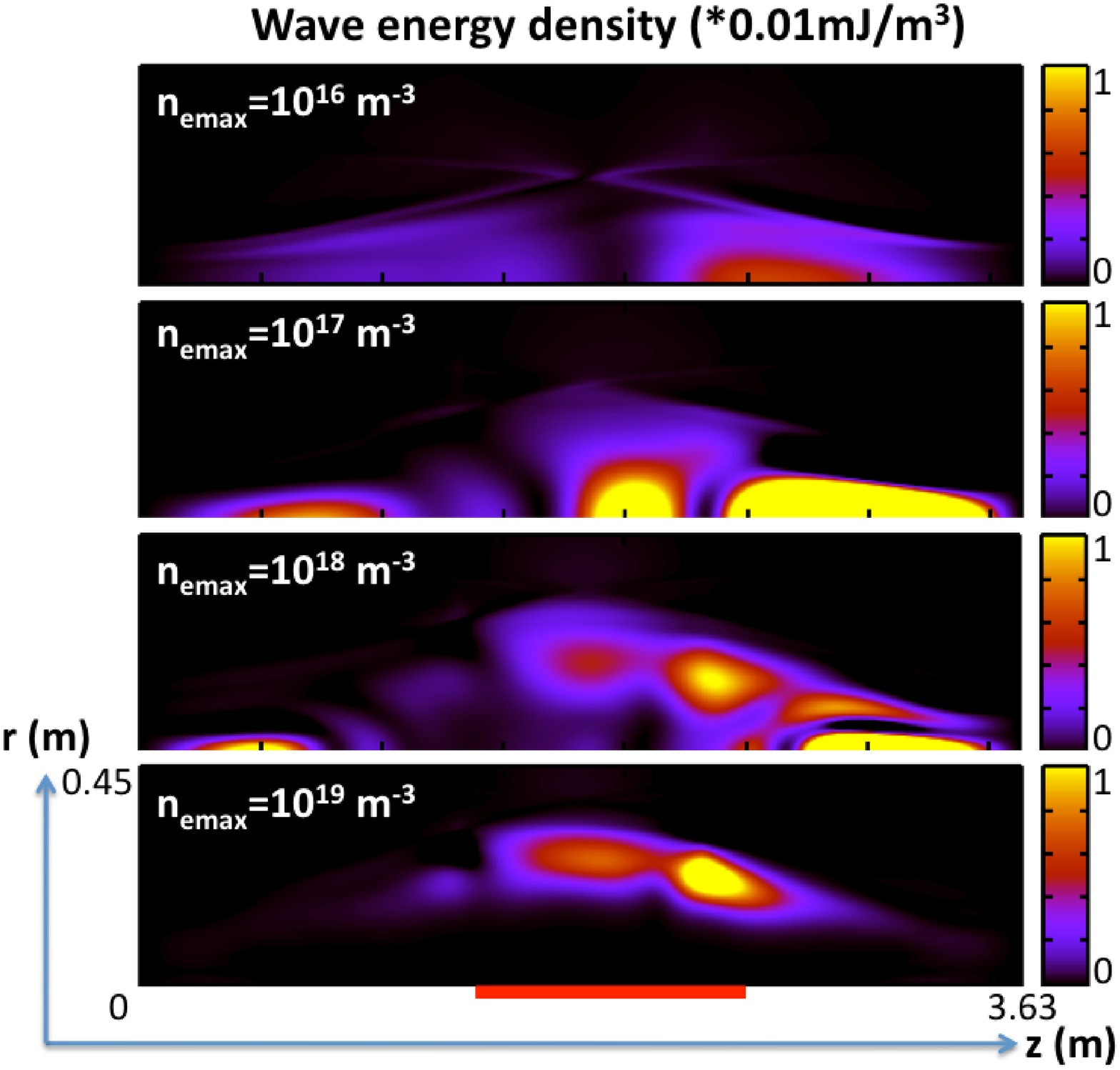}&\includegraphics[width=0.485\textwidth,angle=0]{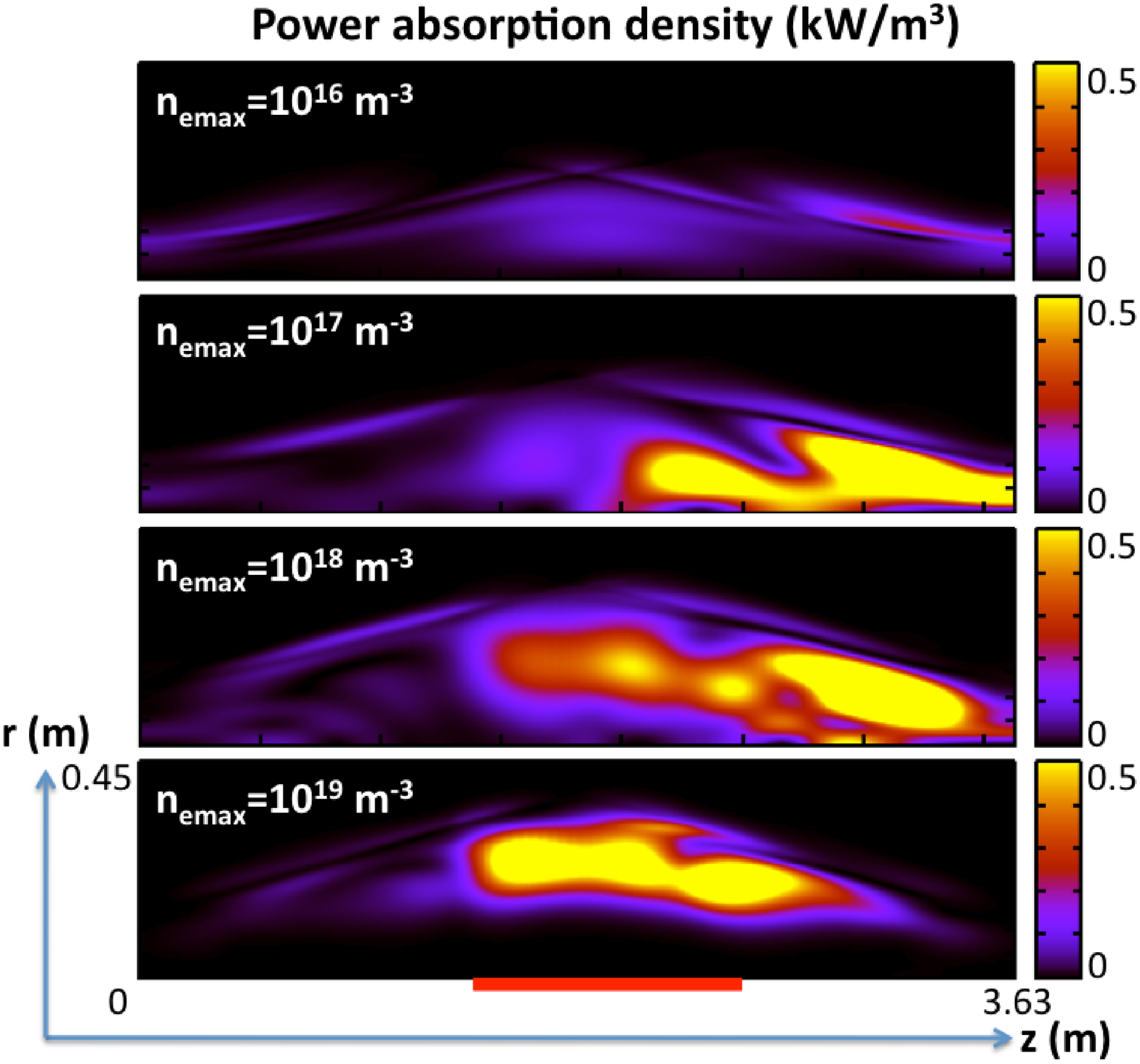}
\end{array}$
\end{center}
\caption{Variations of wave energy density (a) and power absorption density (b) with plasma density for $f=13.46$~MHz and $B_{0max}=0.17~\textrm{T}$. Red bar shows the location of antenna.}
\label{fg6}
\end{figure}

Next, we look into the influence of plasma density on wave activity and power coupling. The frequency is fixed to $13.46$~MHz, close to the mostly used frequency of $13.56$~MHz, and the field strength is also set to $B_{0max}=0.17~\textrm{T}$. Although we vary the maximum plasma density from $n_{emax}=10^{16}~\textrm{m}^{-3}$ to $n_{emax}=10^{19}~\textrm{m}^{-3}$, their normalized radial profiles are kept the same. Figure~\ref{fg5} presents the computed axial profiles of wave field. We can see that the wave magnitude maximizes around $n_{emax}=10^{17}~\textrm{m}^{-3}$. The radial profiles are similar for different plasma densities.
To obtain a full picture, we also show the 2D contour plots of wave energy density and power absorption density in Fig.~\ref{fg6}. It can be seen that both the wave activity and power coupling are strongest for $n_{emax}=10^{17}~\textrm{m}^{-3}$, compared to other density levels. The standing-wave feature in axial direction and shrinking energy distribution are similar to those in Fig.~\ref{fg4}. 

\subsection{Field-strength dependence}\label{fld}
\begin{figure}[ht]
\begin{center}$
\begin{array}{ll}
(a)&(b)\\
\includegraphics[width=0.45\textwidth,angle=0]{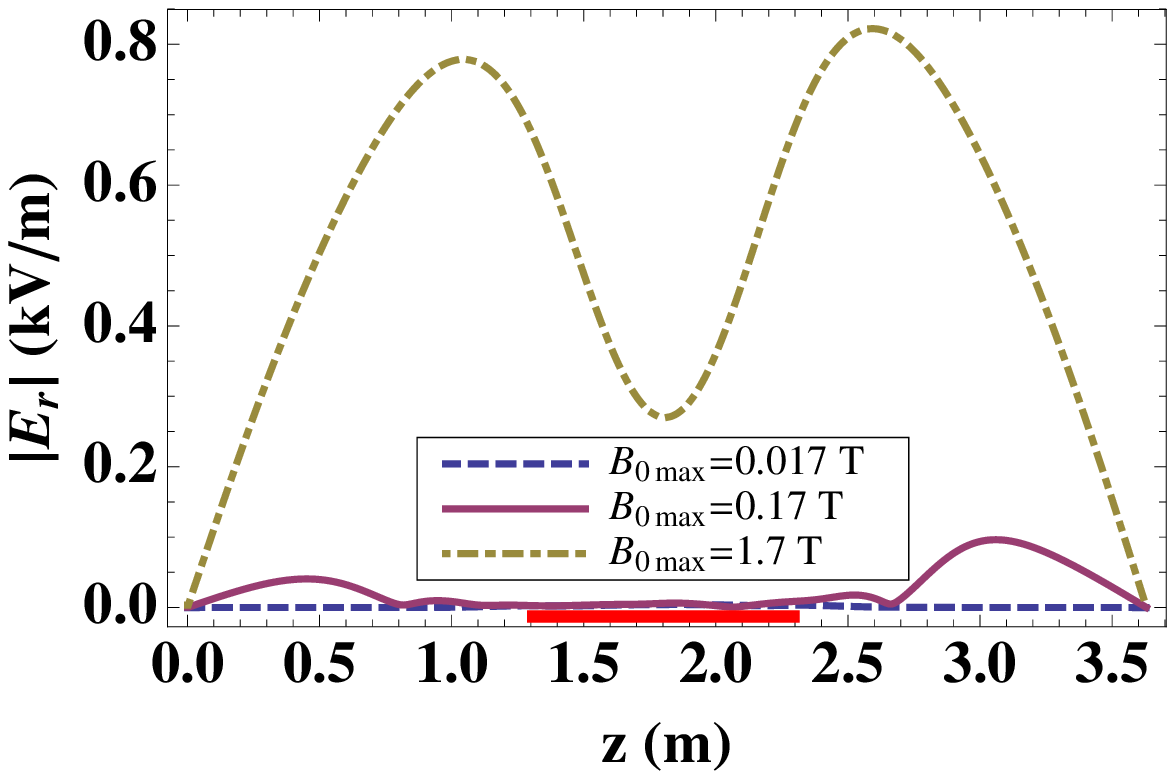}&\includegraphics[width=0.45\textwidth,angle=0]{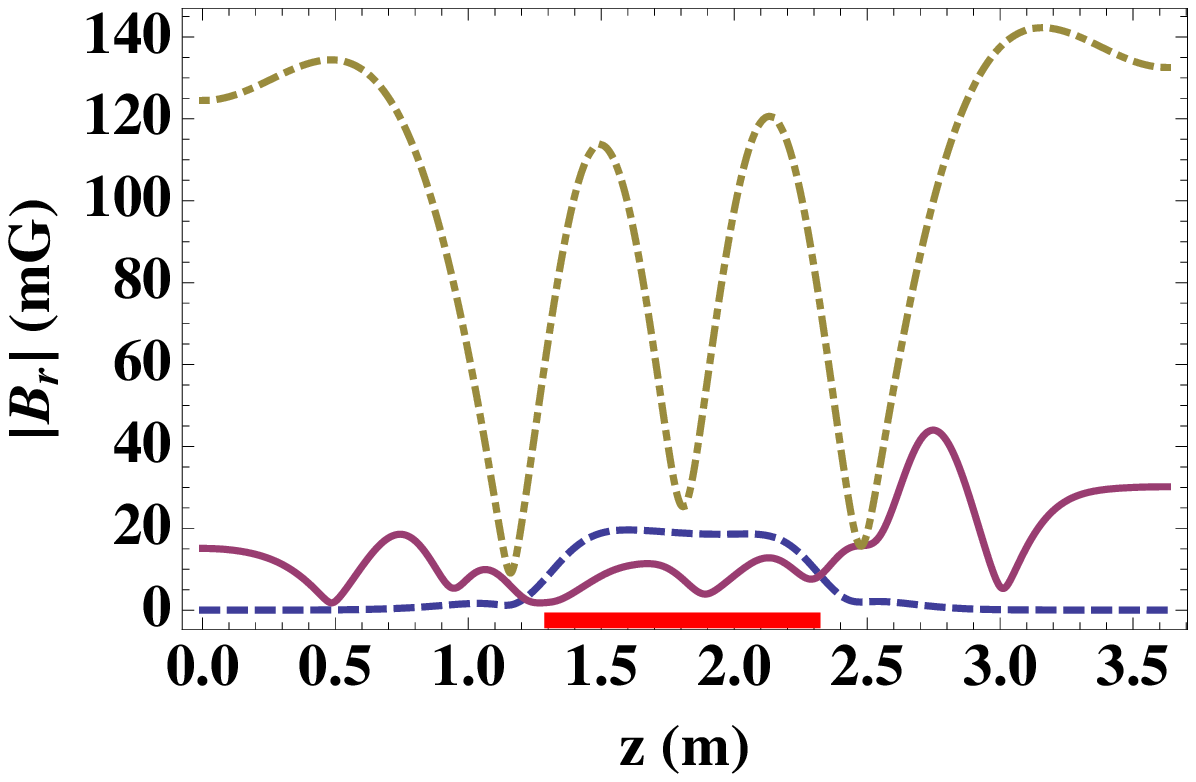}\\
\end{array}$
\end{center}
\caption{Variation of axial wave field profile ($r=0$~m) with equilibrium field strength for $n_{emax}=10^{18}~\textrm{m}^{-3}$ and $f=13.46$~MHz. Red bar shows the antenna location.}
\label{fg7}
\end{figure}

\begin{figure}[ht]
\begin{center}$
\begin{array}{ll}
(a)&(b)\\
\includegraphics[width=0.48\textwidth,angle=0]{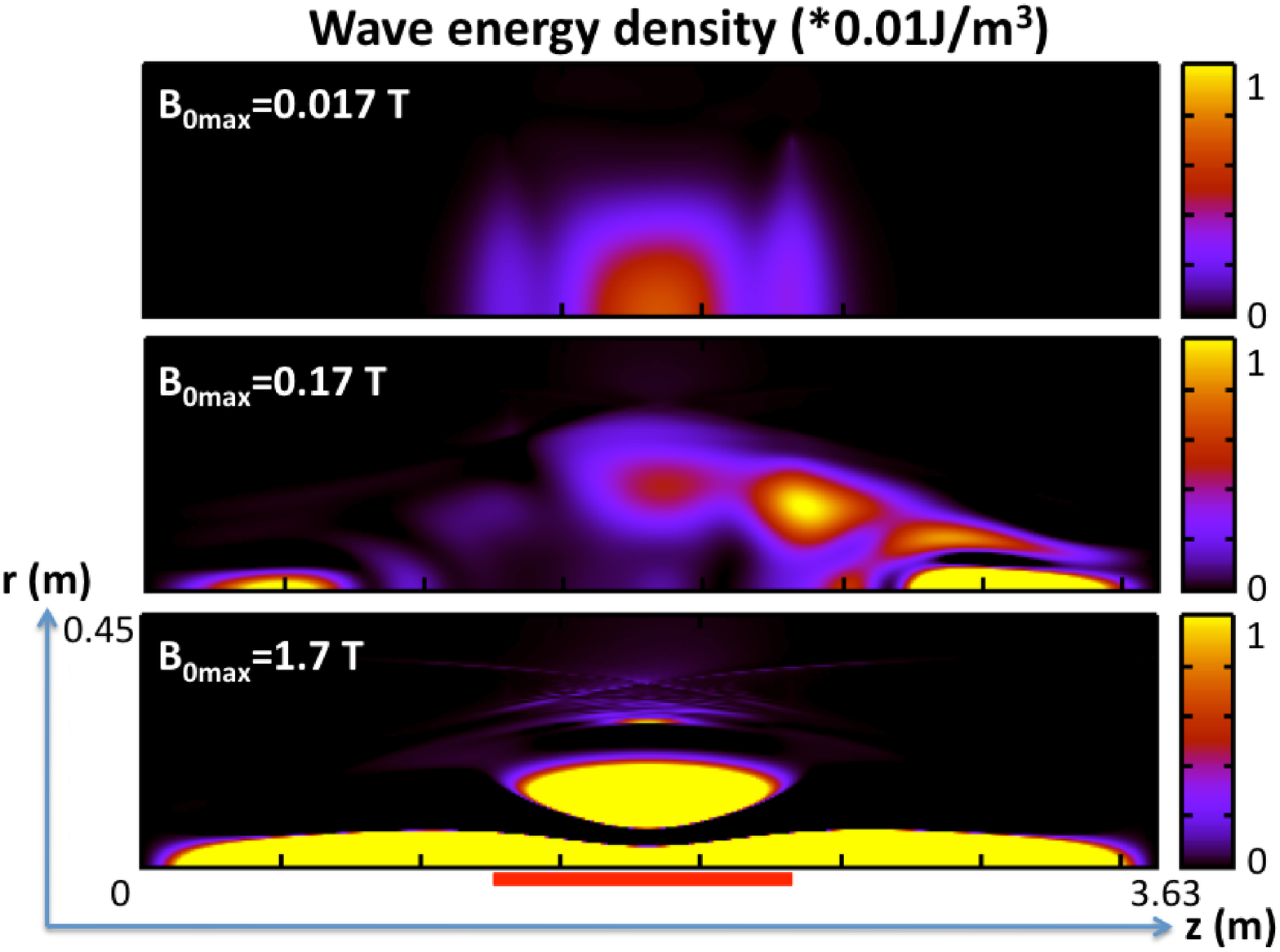}&\includegraphics[width=0.48\textwidth,angle=0]{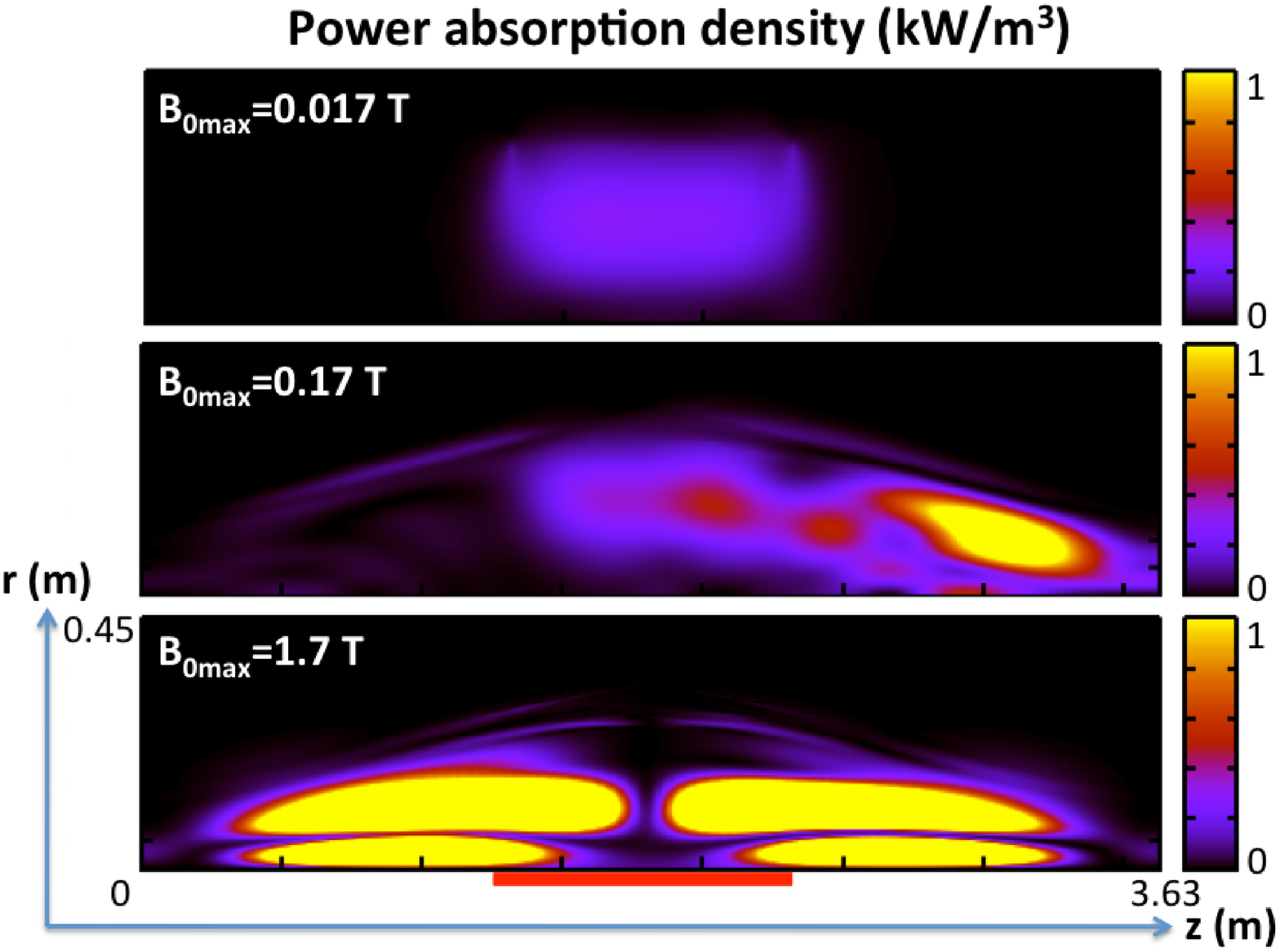}
\end{array}$
\end{center}
\caption{Variations of wave energy density (a) and power absorption density (b) with equilibrium field strength for $n_{emax}=10^{18}~\textrm{m}^{-3}$ and $f=13.46$~MHz. Red bar shows the antenna location.}
\label{fg8}
\end{figure}

Finally, we consider the strength of equilibrium magnetic field. The driving frequency and plasma density are $f=13.46$~MHz and $n_{emax}=10^{18}~\textrm{m}^{-3}$, respectively, and we vary the field strength from $B_{0max}=0.017$~T to $B_{0max}=1.7$~T. Figure~\ref{fg7} shows the computed axial profiles of wave field. We can see that the wave magnitude becomes much stronger when the field strength is higher, and the axial profile shows higher mode number (more periodic structures and shorter wavelength) near the antenna. The radial profiles are largely the same for different field strengths.
The 2D contour plots of wave energy density and power absorption density shown in Fig.~\ref{fg8} also confirm this trend, namely they localize around antenna for low field and spread towards axis and endplates when the field becomes stronger. We can also see axially symmetric distribution regarding antenna and two-peak structure in radius for high field. This is quite different from low field cases, implying that the discharge turns into a different mode. 

Overall, regarding the helicon plasma in a magnetic shuttle, these parameter studies indicate that the power coupling and resulted wave energy are strongest for low driving frequency ($6.68$~MHz), high field strength ($1.7$~T) and plasma density around $n_{emax}=10^{17}~\textrm{m}^{-3}$. This is practically useful for guiding the ongoing HPPX experiments to obtain high-density and large-volume helicon plasma. 

\section{Theoretical analysis}\label{anl}
\begin{figure}[ht]
\begin{center}
\includegraphics[width=0.48\textwidth,angle=0]{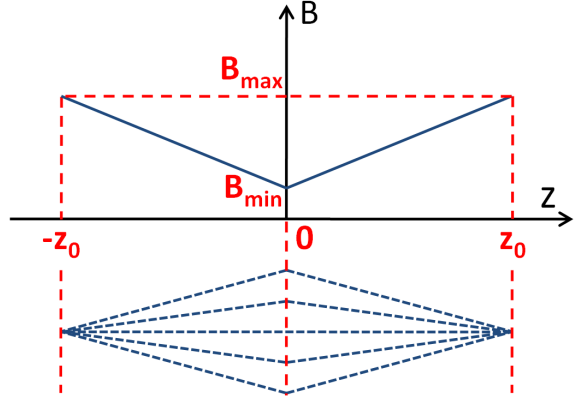}
\end{center}
\caption{A simple example of magnetic shuttle.}
\label{fg9}
\end{figure}

To understand the computed results, we perform theoretical analysis in this section. We simplify the smoothly curved field lines shown in Fig.~\ref{fg2} into straight lines in Fig.~\ref{fg9}, for easy modeling and demonstrating purposes. The solid blue lines show real field strength, while the dashed blue lines illustrate the conceived magnetic shuttle. 
Here we introduce the mirror ratio as $\psi=B_{max}/B_{min}$\cite{Chen:1984aa}. The expression of field strength is then
\begin{equation}\label{eq8}
B(z)=B_{min}\left[1+(\psi-1)\frac{|z|}{z_0}\right].
\end{equation}
The spatial structure of helicon wave field is determined by\cite{Breizman:2000aa}
\begin{equation}\label{eq9}
\frac{1}{r}\frac{\partial}{\partial r}r\frac{\partial E}{\partial r}-\frac{m^2}{r^2}E=-\frac{m}{k^2 r}\frac{\omega^2}{c^2}\frac{E\partial g\partial r}{1+(m\partial g/\partial r)/k^2 r\eta}
\end{equation}
with $E=E_z-(k r/m)E_\theta$ and $c$ the speed of light. For sufficiently dense plasma, which ensures that displacement current is negligible compared to plasma current\cite{Breizman:2000aa}, we could approximate $g=\omega_{p\alpha}^2/\omega\omega_{c\alpha}$. Moreover, considering that electron conductivity along field line is much larger than cross-field Hall conductivity, we ignore the term $(m\partial g/\partial r)/k^2 r\eta$ from the denominator on the right-hand side\cite{Chang:2013aa}, and get
\begin{equation}\label{eq10}
k^2\left(\frac{\partial}{\partial r}r\frac{\partial}{\partial r}r E_\theta-m^2 E_\theta\right)=-E_\theta m \frac{\omega^2}{c^2}\frac{1}{1+(\psi-1)\frac{|z|}{z_0}}r\frac{\partial g_{0}}{\partial r}
\end{equation}
where Eq.~(\ref{eq8}) has been incorporated and $g_0$ is for $B_{min}$. 

We can see that the denominator $1+(\psi-1)(|z|/z_0)$ ranges from $1$ at $z=0$, which is the case for uniform field strength of $B_{min}$, to $\psi$ at $z=z_0$ for the strongest field strength. It implies that the right-hand side of Eq.~\ref{eq10} decreases when steps from midplane to ending throats, and correspondingly the left-hand side has to decrease to satisfy the equation. This yields two consequences: (1) decreased $k_z$ so that the axial wave propagation is slowed or probably reflected, which forms the standing wave features presented in Sec.~\ref{rst}, and (2) reduced radial scale so that wave field and energy distribution are shrunk towards axis (see Fig.~\ref{fg4}, Fig.~\ref{fg6} and Fig.~\ref{fg8}). Please note that these two consequences are general results from magnetic mirror and consistent with precious studies\cite{Stenzel:2018aa, Belov:2007aa}. Moreover, the shrinking feature also agrees with the nature of helicon wave, which is a right-hand circularly polarized mode propagating along magnetic field line\cite{Chen:1984aa, Boswell:1970aa}. Additionally, the inversely proportional relationship between $k$ and $B_0$ can be also found from the simple relation in a slab geometry\cite{Chen:1996ab},
\begin{equation}\label{eq11}
\frac{3.83}{R_p}=\frac{\omega}{k}\frac{n_e e \mu_0}{B_0}
\end{equation}
with $R_p$ the plasma radius and $e$ the electric charge. For fixed driving frequency and plasma density, the axial wave number decreases for increased field strength. 

\section{Summary}\label{sum}
To guide the HPPX experiments to produce high-density and large-volume plasma, this work computes the wave field and power absorption in helicon plasma immersed in a shuttle-shaped magnetic field, using experimental parameters. The definition of magnetic shuttle is first given, namely the magnetic space enclosed by two tandem magnetic mirrors with the same field direction and high mirror ratio. Then the perpendicular structure of wave field along this magnetic shuttle is presented in terms of stream vector plots, which shows significant change from midplane to ending throats, and the vector field rotates and forms a circular layer that separates plasma column radially into core and edge regions near the throats. Through varying the driving frequency, plasma density and the strength of equilibrium magnetic field, we found that the wave magnitude and power absorption decrease for increased driving frequency and reduced field strength, and maximize around $n_{emax}=10^{17}~\textrm{m}^{-3}$ for the conditions employed. This is particularly interesting for the ongoing HPPX experiments. The axial standing-wave feature is always observed, due to the interference between forward and reflected waves from magnetic mirrors, whereas the radial wave field structure stays largely the same. Contour plots of wave energy density and power absorption density all show shrinking feature towards ending throats, which is consistent with the nature of helicon mode propagating along magnetic field lines. Theoretical analysis based on a simple magnetic shuttle and the governing equation of helicon waves shows consistency with computed results and previous studies. 

\ack
This work is supported by Shanghai Engineering Research Center of Space Engine (17DZ2280800) and Chinese Academy of Sciences ``100 Talent" Program (B).

\section{Data Availability Statement}
The data that support the findings of this study are available from the corresponding author upon reasonable request.

\section*{References}
\bibliographystyle{unsrt}

\end{document}